\documentclass[conference]{IEEEtran}
\IEEEoverridecommandlockouts

\usepackage{amssymb}
\usepackage{textcomp}
\usepackage[T1]{fontenc}
\usepackage{mwe}    
\usepackage{float}
\usepackage[bookmarks,bookmarksnumbered]{hyperref}
\hypersetup{colorlinks = true,linkcolor = blue,anchorcolor =red,citecolor = blue,filecolor = red,urlcolor = red,pdfauthor=author}

\usepackage{mathtools}
\usepackage{lineno,hyperref}
\usepackage{algorithmicx}
\usepackage{algorithm}
\usepackage{algpseudocode}
\usepackage{amsmath}
\usepackage{MnSymbol}
\usepackage{wasysym}
\usepackage{graphicx}
\usepackage{csquotes}
\algnewcommand\algorithmicinput{\textbf{INPUT:}}
\algnewcommand\INPUT{\item[\algorithmicinput]}
\algnewcommand\algorithmicoutput{\textbf{OUTPUT:}}
\algnewcommand\OUTPUT{\item[\algorithmicoutput]}
\algnewcommand\algorithmicdefine{\textbf{DEFINE:}}
\algnewcommand\DEFINE{\item[\algorithmicdefine]}
\begin{document}

\title{SHELBRS: Location Based Recommendation Services using Switchable Homomorphic Encryption\\
{\footnotesize \textsuperscript{}}

}

\author{\IEEEauthorblockN {Mishel Jain, Priyanka Singh, Balasubramanian Raman}
\IEEEauthorblockA{
\textit{Dhirubhai Ambani Institute of Information And Communication Technology }\\
Gandhinagar, Gujarat, India \\
\textit{Indian Institute of Technology, Roorkee, Utharakhand, India }\\
Email:{\{201911052, priyanka\_singh\}@daiict.ac.in}, bala@cs.iitr.ac.in}
}

\maketitle

\begin{abstract}
Location-Based Recommendation Services (LBRS) has seen an unprecedented rise in its usage in recent years. LBRS facilitates a user by recommending services based on his location and past preferences. However, leveraging such services comes at a cost of compromising one's sensitive information like their shopping preferences, lodging places, food habits, recently visited places, etc. to the third-party servers. Losing such information could be crucial and threatens one's privacy. Nowadays, the privacy-aware society seeks solutions that can provide such services, with minimized risks. Recently, a few privacy-preserving recommendation services have been proposed that exploit the fully homomorphic encryption (FHE) properties to address the issue. Though, it reduced privacy risks but suffered from heavy computational overheads that ruled out their commercial applications. Here, we propose SHELBRS, a lightweight LBRS that is based on switchable homomorphic encryption (SHE), which will benefit the users as well as the service providers. A SHE exploits both the additive as well as the multiplicative homomorphic properties but with comparatively much lesser processing time as it's FHE counterpart. We evaluate the performance of our proposed scheme with the other state-of-the-art approaches without compromising security. 
\end{abstract}

\section{Introduction}

Location-Based Recommendation Services (LBRS) grant users access to relevant information about their surroundings based on their location and history. For instance, a person searching for a coffee shop nearby his/her location. The service providers would provide the best search results considering his/her present location and previous history. However, availing of such services risks the user's privacy as the shared sensitive information could be misused by these third-party servers to their advantage, causing serious losses to the user \cite{Strava}. This fact kind of delimits the privacy-aware society rushing to leverage such services and creates an urgent need for privacy-preserving recommendation services. 

The real-time location information of the user is handled by location based services(LBS). It also provides recommendation over encrypted history preferences which ensures the user's privacy. Lyu et al. proposed one such state-of-the-art protocol. They adopt the Hilbert curve \cite{sagan2012space} as a mapping tool, collaborative filtering recommender based on the co-occurrence matrix as a recommendation technique \cite{sarwar2001item} \cite{moon2001analysis}, and  Brakerski-Gentry-Vaikuntanathan (BGV) fully homomorphic encryption (FHE) as an encryption scheme \cite{brakerski2014leveled}. However, it is still infeasible for the use of commercial recommendation services due to the high processing time.

In this paper, we propose SHELBRS, a lightweight LBRS that will benefit the users as well as the service providers. Instead of FHE, we employ switchable homomorphic encryption (SHE) that securely switches between partially homomorphic encryption (PHE) schemes. Specifically, Paillier Homomorphic Encryption and ElGamal Homomorphic Encryption for performing additions and multiplications on the encrypted data. PHE evaluates arithmetic operations more efficiently at least 2-3 order of magnitude compared to FHE. SHE supports an arbitrary number of additions and multiplications over encrypted data and serves the principle of FHE with better efficiency. The overall computation and communication cost required in switching between the PHE's is reasonable for real-life applications. 
It overall reduces the processing time without compromising the security.

The remainder of this paper is structured as follows: Section \ref{sec:relatedwork} discusses some of the related works. Section \ref{sec:preliminaries} gives an overview of the Hilbert curve, collaborative filtering based on Co-occurrence Matrix (CM), PHE, and the SHE schemes. Section \ref{sec:LBSR} presents the LBRS using FHE \cite{lyu2019privacy} while Section \ref{sec:proposed technique} details the proposed SHELBRS scheme. Section \ref{sec:results} discusses the experimental results and the  security analysis of the proposed scheme. Section \ref{sec:conclusions} concludes the work along with some future directions.

\section{Related work}
\label{sec:relatedwork}
Lattice-based FHE scheme introduced by Craig Gentry in 2009, is a milestone research that opened doors for proposing possible solutions for encrypted data. It was made possible as this scheme supported computation of arbitrary functions and operations on the ciphertext, without the need of actually decrypting it \cite{gentry2009fully}. 


Many LBS were proposed in the literature to search nearest Point of Interests (POI)'s to the user's private location. In 2003, K-anonymous based technique was introduced which adopts temporal and spatial cloaking \cite{gruteser2003anonymous}. It acquires accuracy but requires a trusted third party to hide the user's location. Private Information Retrieval (PIR) \cite{melchor2008fast}, Private Circular Query Protocol (PCQP) \cite{lien2013novel} and Lightweight Private Circular Query Protocol (LPCQP) \cite{utsunomiya2015lpcqp} are the LBS based on the cryptography methods. In PIR scheme, the user receives POI from the server's database based on Quadratic Residuosity Assumption (QRA) without server's knowledge of which POI a user is interested in. It provides security but takes high execution time for searching POIs. PCQP proposed by Lien et al. is an effective k-NN search algorithm based on paillier cryptosystem and hilbert curve. It secretly shifts the POI-info circularly which is stored on the server. LPCQP, proposed by Utsunomiya et al. is a lightweight protocol that removes unnecessary POI information from the requesting user to reduce computational cost. PIR, PCQP, and LPCQP secure against single point failures and Denial of Service (DOS) attacks. In 2012, Pingley et al. proposed a context-aware scheme for privacy-preserving LBS \cite{pingley2012context}. It projects the user's location on various-grid-length Hilbert curve and uses location perturbation technique to prevent user's privacy from the LBS server. It ensures protection for data privacy. Gang et al. proposed location-based social network (LBSN) towards privacy preservation for "check-in" services in 2019 \cite{sun2019towards}. It designs the framework using k-anonymity based algorithms without using a trusted third-party server. It guarantees secure access and preserves user's location privacy.     

The detail of recommender systems and discussion about different recommendation algorithms based on traditional and network approaches was studied by Lu et al. in 2012.  \cite{lu2012recommender}. It compares the performance of different recommendation algorithms. Badsha et al. introduced an Elgamal cryptosystem based privacy-preserving item-based Collaborative Filtering (CF) in 2016 \cite{badsha2016practical}. It provides recommendations based on the user's average ratings and the similarities between the items. Zhang et al. proposed Factorization Machines(FM) based recommendation algorithm for Rural Tourism in 2018 \cite{zhang2018fm}. It provides recommendations based on geographical distribution and seasonal features such as the user's best suitable season for traveling, how many kilometers is the traveling spot away from the city, and other user's reviews or comments for the particular location. In 2018, Horowitz et al. proposed a mobile recommender system named "EventAware" for the events \cite{horowitz2018eventaware}. It provides recommendations on the basis of both context-aware and tag-based algorithms. In 2019, Qi et al. proposed a time-aware and privacy-preserving distributed recommendation service based on a locality-sensitive hashing (LSH) to provide most accurate recommended results \cite{qi2019time}. Papakyriakopoulos et al. analyzed the political networks based on hybrid collaborative filtering and deep learning recommender algorithms in 2020 \cite{papakyriakopoulos2020political}. It shows how hyperactive users influence recommendations. It also compares the results based on likes and comments with and without the inclusion of the hyperactive users along with the rest of the users in the datasets. 

Combining both location privacy and privacy-preserving recommendations, Lyu et al. proposed privacy-preserving recommendations for LBS in 2019. It provides suggestions over encrypted previous histories considering user's live location data. This protocol uses one trusted third party server where sensitive data such as crypto keys are stored and one honest but curious server where all the computations are performed.  

\section{Preliminaries}
\label{sec:preliminaries}
\subsection{Hilbert Curve}
Hilbert curve is a mapping tool that is used to transform 2-D space into 1-D space. It preserves the adjacency of the neighboring points and has best clustering properties. In order to preserve the adjacency property, the orientation is not retained \cite{sagan2012space} \cite{moon2001analysis}. As we increase the order of a pseudo-Hilbert curve, a given point on the line converges to a specific point. The two data points which are close to each other in 2-D space are also close to each other after mapping into 1-D space. 

\begin{figure}[ht]
\centering
\includegraphics[scale=0.55]{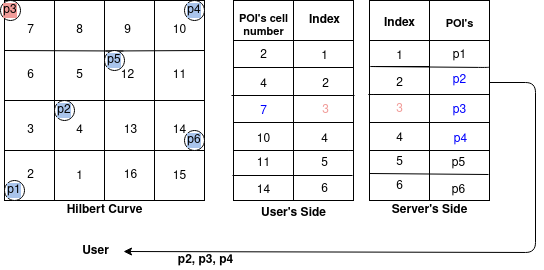}
\caption{The combination of Continuous Hilbert space-filling curve and Location-based services}
\label{fig:hilbert}
\end{figure}

Fig. \ref{fig:hilbert} shows a Hilbert curve over the landscape, storing the POI look-up table at the user end and the POI information table at the server end. The user $p3$ is located on  cell 7 and he/she is requesting POIs from the server. At the user's side, p3 is located at Index-3 and after sending index information to the server, the server sends nearby POIs i.e. p2, p3, p4 corresponding to the current index back to the user. This is how the nearest POIs to the user's location is suggested when the landscape is mapped on the Hilbert curve. However, sending such information in plaintext does not ensure the user's privacy.

\subsection{Collaborative filtering (CF) Recommender based on Co-occurrence Matrix (CM)}
CF Recommender consists of two well-known algorithms: user-based CF and item-based CF. We used an item-based CF recommender as similarities between items are more stable than that of users. It finds the similarity between items and provides the best recommendation. Some E-commerce websites such as Amazon provides recommendation such as "A person bought product A also bought product B" or "A person liked the cafe A also liked cafe B". 

CM contains the visited POIs information. It computes the number of times each pair of items occurs together in the user-item inversion list. To generate CM, the first step is to generate a user-item inversion list and the second step is to traverse the list and follow the algorithm as described: 
\begin{itemize}
    \item $CM[i][j] (i != j)$ is increased by 1 if item $i$ and the item $j$ are in the same user's inversion list.
    \item $CM[i][j] (i == j)$ is increased by $1$ for every item $i$.
\end{itemize}

\begin{figure}[ht]
\centering
\includegraphics[scale=0.65]{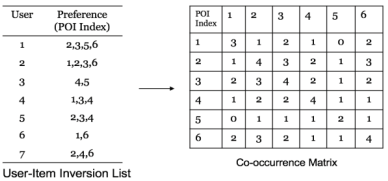}
\caption{Formation of the final CM \cite{lyu2019privacy}}
\label{fig:CM}
\end{figure}

Fig. \ref{fig:CM} shows the formation of CM based on the user-item inversion list. According to the user-item inversion list, $User_{1}$ has the preference for indices $2$, $3$, $5$, $6$. The value at all possible pairs of indices in $CM$ such as $CM[2][3]$, $CM[2][5]$, $CM[2][6]$, $CM[3][2]$, $CM[3][5]$, $CM[3][6]$, $CM[5][2]$, $CM[5][3]$, $CM[5][6]$, $CM[6][2]$, $CM[6][3]$, $CM[6][5]$ is incremented by $1$. Every time an item occurs in the list, the value in $CM$ such as $CM[2][2]$, $CM[3][3]$, $CM[5][5]$, and $CM[6][6]$ is incremented by $1$. Likewise for all the users, the above algorithm is performed to get the final $CM$.

\subsection{Partially homomorphic Encryption (PHE)}
\label{subsec:PHE}
PHE schemes are a kind of encryption schemes that allow only certain types of operations on the encrypted data. If we decrypt the processed encrypted data, the results would be the same as if calculated over the corresponding plaintext values. Based on the type of operations supported, it can be categorized as additive PHE or multiplicative PHE. For instance, Paillier is an example of additive PHE and ElGamal, an example of multiplicative PHE. We will briefly describe each of them along with their homomorphic properties.  

\textbf{Paillier Encryption as additive PHE} :
\begin{itemize}
    \item \textbf{KeyGen($ 1^{n} , + $):} On input a security parameter $1^{n}$, the algorithm chooses $(N, p, q)$ where $ N = p * q $,  $p$ and $q$ are $n$ bit primes, and $\phi(N) = (p-1)* (q-1) $. The Paillier ADD scheme public-private key pair:
    \begin{equation}
       \left \langle pk^{+}, sk^{+}  \right \rangle = \left \langle N,(N, \phi(N))\right \rangle
    \end{equation}
    
    \item \textbf{Enc($pk^{+} , m$):} The algorithm takes a plaintext $m$ and a public key $N$ as input. It chooses a random $r$ $\in\, \,Z_{N}^{*}$ and outputs the ciphertext:
    
    \begin{equation}
        c^{+}=(1+N)^{m}.(r)^{N}\mod N^{2}       
   \end{equation}
   
   \item \textbf{Dec($sk^{+} , c^{+}$):} The algorithm takes a ciphertext $c^{+}$ and a private key $(N, \phi(N))$ as input and outputs the message:
   
   \begin{equation}
        m = \dfrac{((c^{+})^{\phi(N)} \mod N^{2}) - 1}{N} . \phi(N)^{-1} \mod N
   \end{equation}
   \\
   \item \textbf{Paillier homomorphic properties}:
   \begin{enumerate}
       \item \textbf{Addition}: The product of two encrypted ciphertexts results in the sum or addition of their corresponding plaintexts: 
       
        \begin{equation}
            \label{eq:e20}
            \footnotesize
            Dec(E^{+}(m_{1})*E^{+}(m_{2})\mod N^{2})=(m_{1} + m_{2})\mod N 
        \end{equation}
      
       \item \textbf{Scalar  Multiplication}: Raising a scalar to the power of encrypted ciphertext results in the product of the scalar and the corresponding plaintext:
       \begin{equation}
            Dec(E^{+}(m_{1})^{k} )\mod N^{2}) = (k * m_{1})\mod N   
       \end{equation}
       
       
   \end{enumerate}
   
\end{itemize}

\textbf{ElGamal Encryption as multiplicative PHE}: 
\begin{itemize}
    \item \textbf{KeyGen($ 1^{n}, * $):} On input a security parameter $1^{n}$, the algorithm chooses $(N, p, q)$ where $ N = p * q $, $p$ and $q$ are $n$ bit primes, considers $g$ as square value and sets $ g = 16 $.  It also chooses a random odd number $x$, and sets $h = g^{x} \mod N$. The ElGamal MUL scheme public-private key pair:
    \begin{equation}
       \left \langle pk^{*}, sk^{*}  \right \rangle = \left \langle (N, g, h),(N, g, x)\right \rangle
    \end{equation}
    
    \item \textbf{Enc($pk^{*}, m$):} The algorithm takes a plaintext $m$ and a public key $(N, g, h)$ as input. It chooses a random $r$ $\in\, \,Z_{N}^{*}$ and outputs the ciphertext:
    
    \begin{equation}\label{eq:e8}
        c^{*}=\left \langle c_{1}^{*}, c_{2}^{*}  \right \rangle  
        =\left \langle mh^{r}, g^{r} \mod N  \right \rangle  
   \end{equation}
   
   \item \textbf{Dec($sk^{*}, c^{*}$):} The algorithm takes a ciphertext $c^{*}$ and a private key $(N, g , x)$ as input and outputs the message:
   
   \begin{equation}
        m = \dfrac{c_{1}^{*}}{(c_{2}^{*})^{x}} \mod N
   \end{equation}
   
   \item \textbf{ElGamal homomorphic properties}:
   \begin{enumerate}
       \item \textbf{Multiplication}: The product of two encrypted ciphertexts results in the multiplication of their corresponding plaintexts: 
       
        \begin{equation}
            \footnotesize
            Dec(E^{*}(m_{1})*E^{*}(m_{2})\mod N^{2})=(m_{1} * m_{2})\mod N 
        \end{equation}
    \end{enumerate}  
    
\end{itemize}

\subsection{Switchable homomorphic Encryption (SHE)}
\label{subsec:SHE algo}
We work with a variant of ElGamal MUL scheme $E^{*}$ and Paillier ADD scheme $E^{+}$. ElGamal MUL scheme uses a large composite modulus i.e., $N = p * q $, where $p$ and $q$ are large primes. Both the partially homomorphic schemes share the same modulus. We consider two servers, say a server and a proxy.

The algorithms used in the SHE scheme are described as follows:
\begin{itemize}
    \item \textbf{KeyGen($ 1^{n} $)}: On input a security parameter $1^{n}$, the algorithm outputs $(N, p, q)$, where $ N = p * q $, $p$ and $q$ are $n$ bit primes, and $\phi(N) = (p-1)(q-1)$. The Paillier ADD scheme public-private key pair:
    \begin{equation}\label{eq:e3}
       \left \langle pk^{+}, sk^{+}  \right \rangle = \left \langle N,(N, \phi(N), p, q)\right \rangle
    \end{equation}
    It also chooses the generator  $g = 16$, two random odd numbers $x_{0},x_{1} \in\, \,Z_{N}^{*}$ where
    $\left | x_{0} \right | \approx \left | x_{1} \right |  < 1/2\left | N \right |$. It sets $x=x_{0}x_{1}$ and $h=g^{x}$. The ElGamal MUL scheme public-private key pair:
   
   \begin{equation}\label{eq:eq11}
        \left \langle pk^{*},sk^{*} \right \rangle = \left \langle (N,g,h),(N,g,x_{0},x_{1}) \right \rangle      
   \end{equation}
   
   \item \textbf{Enc($pk^{o}$, $m$)}: The algorithm runs $E^{+}$ encryption scheme if o is '+' else runs $E^{*}$ encryption scheme.
   
   \item \textbf{Dec($sk^{o},c^{o}$)}: The algorithm runs $E^{+}$ decryption scheme if o is '+' else runs $E^{*}$ decryption scheme.
   
   \item \textbf{KeyShaGen($sk^{+}$)}: The algorithm sets both the secret key shares $k_{0}^{+}$ (proxy) and $k_{1}^{+}$ (server) to NULL.
   
   \item \textbf{KeyShaGen($sk^{*}$)}: The algorithm sets both the secret key shares $k_{0}^{*}$ (proxy) and $k_{1}^{*}$ (server) to $x_{0}$ and $x_{1}$ respectively.
   
   \item\textbf{AddToMul($c^{+},pk^{*}$)}: The algorithm is run locally by the server. Given an ADD ciphertext of the form:
   
   \begin{equation}
        E^{+}(m)=(1+N)^{m}.(r')^{N}\mod N^{2}       
   \end{equation}
   
   and the MUL public key $pk^{*} = (N,g,h)$, the algorithm chooses a random $r$ $\in\, \,Z_{N}^{*}$ and outputs the encrypted MUL ciphertext:
   
   \begin{equation}\label{Eq:e1}
       E^{+}(E^{*}(m))=\left \langle (1+N)^{mh^{r}}.(r')^{Nh^{r}}\mod N^{2}
       ,g^{r} \right \rangle
   \end{equation}
   
   \item\textbf{MulToAdd($c^{+},k_{0}^{*},k_{1}^{*}$)}: The algorithm is jointly run by a server and a proxy. On input an encrypted MUL ciphertext of the form (\ref{Eq:e1})
    the server chooses a random $s$ $\in\, \,Z_{N}^{*}$, computes
    
    \begin{equation}\label{Eq:e2}
        c'= (g^{r+s})^{k_{1}^{*}},
        R = g^{s}     
    \end{equation}
    
    and forwards (\ref{Eq:e1}) and (\ref{Eq:e2}) to the proxy.
    
    The proxy then using its key shares $k_{0}^{*}$, computes $(c')^{k_{0}^{*}}=h^{r+s}$ and finds its inverse $(h^{r+s})^{-1}$. It also computes and returns 
    \begin{align}
        c'' &= ((1+N)^{m{h^r}}.(r')^{N{h^r}})^{{h^{r+s}}^{-1}} \mod N^{2}\nonumber\\
            &= (1+N)^{m{h^{-s}}}.(r')^{N{h^{-s}}} \mod N^{2}\\
            &= E^{+}(mh^{-s})\nonumber    
    \end{align}

    and $R':=R^{k_{0}^{*}}$ to the server.
    
    Finally, the server computes $(R')^{k_{1}^{*}}=h^{s}$ and recovers the corresponding ADD ciphertext $E^{+}(m)$ by homomorphically removing $h^{-s}$ from $c''$.
\end{itemize}

\section{Lyu et al.'s Protocol}
\label{sec:LBSR}

In this section, we give an overview of Lyu et al. protocol which was meant to recommend services based on the current location of a user and his past behavior without compromising his privacy \cite{lyu2019privacy}. 
It solved the problems existing in the state-of-the-art privacy-preserving algorithms that were based on k-NN technique for searching POI's \cite{lien2013novel}. The major bottleneck was that the recommendation service didn't consider the user's past behavior while recommending any services that ultimately resulted in failing to attract the user's usage of the recommendation system. Another major demerit was it lacked any benefits for the service providers facilitating such services as the private keys were available only to the user and hence, the service providers could not extract any information from the user's data towards making their profits.

Lyu et al. resolved the aforementioned issues using collaborative filtering technique that works on top of database encrypted using FHE, besides encrypting the user's location and preferences. Also, it allowed the service providers to extract some aggregate information based on the user's data via an introduction of a Privacy Service Provider (PSP) that generates and holds the private keys. This increased the commercial value of the recommendation service but still the need of heavy computational resources required for FHE restricts the usage.\\


\begin{figure*}[ht]
\centering
\includegraphics[scale=0.70]{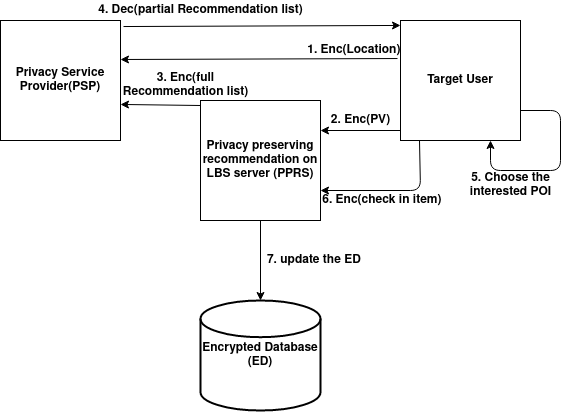}
\caption{An overview of Lyu et al.'s protocol}
\label{fig:overview}
\end{figure*}

\subsection{System Model}
An overview of the protocol is shown in Fig. \ref{fig:overview}. It involves three main components: 
\begin{itemize}
\item \textbf{Privacy-Preserving Recommendation Server (PPRS):}  PPRS is a semi-trusted i.e honest but curious entity that is responsible for finding the nearest POIs to the user, by considering the similarity between the POIs near the user and his current location. It performs  two main tasks: The first task is to calculate the recommendation list based on preference vector $PV$ and co-occurrence matrix $CM$. Here, $PV$ provides a rating of a user for a particular item and $CM$ describes the similarities between the items. The second task is to calculate the aggregated user behavior over the encrypted database ($ED$).

\item \textbf{Privacy Service Provider (PSP):} PSP is a trusted third party which makes a profit from user's behavior statistics. It holds and generates the private and public key pairs. It is responsible for providing public keys to users, ED and PPRS whenever the requests arrive. It generates partial recommendation list based on user's location information. 
    
\item \textbf{Encrypted Database (ED):} It stores $CM$ encrypted by FHE. 
\end{itemize}

\subsection{Description of Lyu et al.'s Protocol}
This section describes Lyu et al.'s protocol.  First, we describe briefly the three main phases of the protocol and then go for a detailed step-by-step description. The three main phases are as follows: \\

\textbf{Initialization Phase:}
 Personal co-occurrence matrix ($CM_{u}$), which contains the information of visited POIs, is generated by the user on the basis of his/her preferences. Each user sends his/her encrypted $CM_{u}$ to the PPRS. PPRS constructs the final $CM$ by combining these $CM_{u}'s$. The operation performed here is exploiting the homomorphic addition property of FHE operation. This combined matrix $CM$ is stored in the $ED$.\\

\textbf{Recommendation Phase:}
For computing full recommendation list for the user, each item's prediction value is computed by performing homomorphic addition and multiplication property of FHE. Prediction value is derived as
   \begin{equation}\label{eq:e21}
    P_{u.i}=\sum_{j \in N(u)} (w_{ij}*r_{uj})
    \end{equation}
where,

$N(u)$ denotes all the items, $w_{ij}$ is the similarity between item $i$ and $j$, and $r_{uj}$ is the rating of user u for item j. \\

\textbf{ED Updation Phase:}
ED stores the $CM$ and it needs to be updated according to the user's new behaviors. The updation occurs as follows: 

\begin{itemize}
    \item After receiving recommendation results from PSP in plain text, the target user selects any POI of his/her choice.
    \item He/She then sends the encrypted results to the ED for further updated recommendation.
    \item Instead of sending the whole $CM$, each user sends only the difference from its original $CM$ to update the matrix with the latest information. It protects the privacy of the user from PPRS by sending the matrix in an encrypted FHE domain.
\end{itemize}

The detailed step-by-step description of the Lyu et al.'s protocol is as follows:\\ 
\\
\textbf{Step 1:} Initially, the public key ($p_{k}$) is distributed by the PSP to a user and PPRS. The target user sends her/his encrypted location to PSP. \\
\textbf{Step 2:} The target user also sends encrypted preference vector to PPRS at the same time.\\
\textbf{Step 3:} PPRS generates and sends the encrypted full recommendation list to PSP. \\
\textbf{Step 4:} After PSP decrypts both the target user’s location and the full recommendation list, it scans the whole list and generates the partial recommendation list according to the user's location information. It ensures user's privacy by sending it through an encrypted channel to the target user.\\
\textbf{Step 5:} The target user selects the POI from the partial recommendation list.\\
\textbf{Step 6:} She/he sends the encrypted result to the PPRS. \\
\textbf{Step 7:} PPRS then updates the ED for providing the best recommendation.

\section{Proposed SHELBRS Protocol}

In our proposed framework, we replace the FHE component with SHE to minimize the overall computational complexity and also, speed up the entire process so that it could be better suited for real-life scenarios. The security of the proposed protocol is kept at par with the corresponding FHE based protocol. Our protocol does not make use of any trusted third party server to store crypto keys. It simply sends the secret key shares between the servers so that an individual server cannot leak any user's sensitive information. 
\label{sec:proposed technique}

\begin{figure}[ht]
\centering
\includegraphics[scale=0.50]{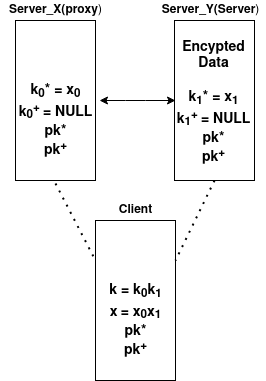}
\caption{Secure computation via two servers}
\label{fig:recommend}
\end{figure}


The details of each stage of SHELBRS protocol is discussed as follows:  

\subsection{Setup Stage}
The setup of SHELBRS is based on a client-server architecture. We consider two servers, say server $X$ and server $Y$, and the interaction between the servers or a server and a client is shown in Fig. \ref{fig:recommend}. Security holds as long as atleast one of the servers is honest i.e. they do not collude by sharing cryptographic keys. Let us assume that the landscape $I$ is mapped on a Hilbert curve and divided into indices $I_{1},I_{2}, \ldots I_{n}$. Based on a client-server model, a client is located on one of the indices and has its own preference vector. The encrypted database of $CM$ is already stored at server $Y$. A client generates a public-private key pairs using (\ref{eq:e3}) (\ref{eq:eq11}) and sends public keys $pk^{+}$ and $pk^{*}$ to both the servers. Clients uses KeyShaGen($sk^{*}$) and KeyShaGen($sk^{+}$) algorithms as described in Section \ref{subsec:SHE algo}  and sends $k_{0}^{*}$ and $k_{0}^{+}$ to server X and $k_{1}^{*}$ and $k_{1}^{+}$ to server $Y$. The POIs nearby user's location is recommended by the computations which are being performed on the servers.

\subsection{Initialization Stage}
This stage describes the steps which needs to be computed before the client starts executing his/her role.

\begin{figure*}[!ht]
\centering
\includegraphics[scale=0.70]{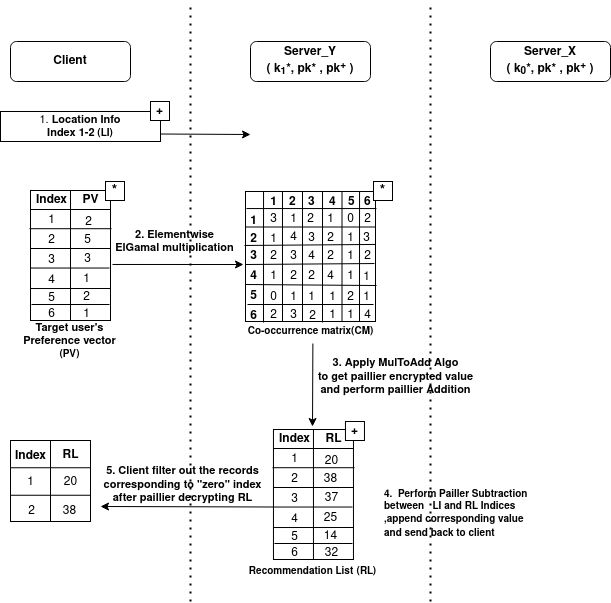}
\caption{SHELBRS: Proposed Recommendation}
\label{fig:detailedrecommend}
\end{figure*}

\begin{itemize}
    \item  Personal co-occurrence matrix ($CM_{u}$) contains the information of visited POIs. During the initialization stage, each user generates his/her personal $CM_{u}$ based on initial users’ preference. 
    \item An user $u$ sends $CM_{u}$, which is encrypted by the public key $pk^{+}$, to the server $Y$.
    \item The end task is to merge all $CM_{u}$ to generate the final $CM$. This requires paillier homomorphic Addition property as described in (\ref{eq:e20}) and AddToMul algorithm as discussed in Section \ref{subsec:SHE algo}.
    
    \item The encrypted CM is stored at server Y.
\end{itemize}

\begin{algorithm}[ht]
    \renewcommand{\algorithmicrequire}{\textbf{Input:}}
    \renewcommand{\algorithmicensure}{\textbf{Output:}}
    \renewcommand{\algorithmicdefine}{\textbf{Define:}}
    \caption{Recommendation}
     \begin{algorithmic}[1]
     \INPUT{$CM^{*}$, $PV^{*}$, Item Index set $I^{*}$}
     \OUTPUT{Recommendation list $RL$ for all items } 
     
     \DEFINE \begin{enumerate}
        \item {$CM^{*}$ is ElGamal encrypted co\_occurrence matrix}
        \item {$PV^{*}$ is ElGamal encrypted history preference vector}
        \item {RL[i] is item $i's$ recommendation score}
    \end{enumerate}
    \Procedure{recommend}{$CM^{*}$ , $PV^{*}$ , $I^{*}$, size}
    \State{Assign $RL^{+}[1 \cdots size] = 0$}
    \For{$i = 1;$\ i <= size;\ $i = i + 1$}
        \For{$j = 1 ;\ j <= size ;\ j = j + 1$}
        \begin{align}
            ctotal &= \textbf{elGamalMultiplication}( CM[i][j]^{*} , PV[j]^{*})\nonumber\\
            c1 &= \textbf{pallierEncryption}(pkadd, ctotal[0])\nonumber\\
            temp &= \textbf{MulToAdd}( [c1, ctotal[1]], k_{1}^{*}, k_{0}^{*} )\nonumber\\
            RL[i]^{+} &= \textbf{paillierAddition}( RL[i]^{+} , temp )\nonumber    
        \end{align}
             
        \EndFor
    \EndFor
    \State {\textbf{return} $RL^{+}$}
    \EndProcedure
\end{algorithmic}
\end{algorithm}

\subsection{Protocol Operation stage}
The detailed process of how the recommendation is being generated is shown in Fig. \ref{fig:detailedrecommend}. The client and the servers interact in the following manner:

\textbf{Step 1:} The client encrypts history preference vector $PV$ using $Enc(pk^{*},PV)$ and location info using $Enc(pk^{*}, Location\_info)$ and sends it to server $Y$.\\

\textbf{Step 2:} The client computes each item's prediction value $P$ using (\ref{eq:e21}) to generate a recommendation list. The prediction is generated using Algorithm 1. 

 To calculate each index's recommendation score, ElGamal encrypted $CM^{*}$ and $PV^{*}$ are elementwise multiplied using the multiplicative property of \textbf{ElGamal Encryption}. The corresponding result is transformed into encrypted ElGamal ciphertext  using \textbf{Paillier Encryption} scheme in Section \ref{subsec:PHE}.
  
  \textbf{Step 3:} It is further converted into paillier encrypted ciphertext using \textbf{MulToAdd} algorithm in Section \ref{subsec:SHE algo}. The corresponding result is finally added using additive property of \textbf{Paillier Encryption} to generate the corresponding recommendation score. 
  
  Likewise, the above steps 2 and 3 are executed for each index to generate the final recommendation list.  
 
    \textbf{Step 4:} According to the target user’s location, the final recommendation list is filtered out.
    
    The server $Y$ performs paillier homomorphic subtraction property corresponding to the indices stored in the recommendation list and user's location. It appends the result to the recommendation list and the updated list is sent to the client. 
    
    \textbf{Step 5:} The client then decrypts it using paillier decryption algorithm and filters out the records corresponding to the value `0'.
    The client chooses one of the locations(indices) from the recommendation list according to his/her choice and update his/her behavior in the inversion list. Each client sends only his/her new paillier encrypted $CM_{u}$ to server Y which is the difference between the current $CM_{u}$ and the client's original $CM_{u}$ before the recommendations.

    
\section{Experimental Results}
\label{sec:results}
To validate the proposed protocol based on SHE, the experiment is executed on Ubuntu 20.04.2 LTS powered by Intel® Core™ i5-6200U CPU @ 2.30GHz × 4 processor and RAM 8 GB. We have considered a client and two servers on the same machine. In this experiment, we considered the artificial dataset with POIs in range \{10, 20, 40, 80, 100, 1000, 2000, 3000, 4000, 5000\}.


The $CM$ is already stored on the server $Y$ and then, the client starts executing his/her behavior. So, the execution time taken for the computations during initialization phase to generate $CM$ is not considered in the total computation cost taken by the client. The time taken to generate public-private keypair is constant. The updation of $ED$ can be performed even when a user is offline, so it does not affect the efficiency of the system.

\begin{table}[ht]
    \centering
    \begin{tabular}{|p{2.0cm}|p{1.5cm}|p{2.2cm}|p{1.5cm}|}
        \hline
        \textbf{Total Elements} & \textbf{Encryption Time[s]} & 
        \textbf{Recommendation Time[s]} & \textbf{Decryption Time[s]}\\
        \hline
        10 & 0.001 & 0.022  & 0.001\\
        \hline
        20 & 0.001 & 0.081 & 0.001\\
        \hline
        40 & 0.001 & 0.299 & 0.002\\
        \hline
        80 & 0.003 & 1.142 & 0.004\\
        \hline
        100 & 0.004 & 1.923 & 0.006\\
        \hline
        1000 & 0.04 & 197.99 & 0.058\\ 
        \hline
        2000 & 0.066 & 732.107 & 0.104\\ 
        \hline
        3000 & 0.138 & 1668.944	& 0.157\\ 
        \hline
        4000 & 0.164 & 2888.376 & 0.232\\ 
        \hline
        5000 & 0.17 & 4355.451 & 0.3\\ 
        \hline
        
    \end{tabular}
    \caption{Computation Cost}
    \label{Recommendation time}
\end{table}
The total computation cost involves encryption of a client's preference vector, computation of recommendation list and decryption of recommendation list. In Table~\ref{Recommendation time}, we measured encryption time, recommendation time,and decryption time for the indices in an encrypted domain. We have run the experiment five times for each index and taken an average of it. 

The experiment uses encryption which adds extra computation cost over plaintext. So, we calculated and compared the total computation cost for the plaintext and encrypted domain up to 1000 indices as shown in Table~\ref{plainencrypted}. 
\begin{table}[ht]
    \centering
    \begin{tabular}{|p{2.0cm}|p{2.0cm}|p{2.0cm}|}
        \hline
        \textbf{Total Elements} & \textbf{Plaintext Domain} & 
        \textbf{Encrypted Domain}\\
        \hline
        10 & 0.001 & 0.024\\
        \hline
        20 & 0.001 & 0.083\\
        \hline
        40 & 0.002 & 0.302\\
        \hline
        80 & 0.005 & 1.149\\
        \hline
        100 & 0.007 & 1.933\\
        \hline
        1000 & 1.32 & 198.088\\ 
        \hline
    \end{tabular}
    \caption{Comparison of total execution time in plaintext domain and encrypted domain}
    \label{plainencrypted}
\end{table}

We also plotted the graph comparing the total execution time taken by SHELBRS scheme and Lyu et al. protocol in Table~\ref{ComparingComputationCost}. 

\begin{table}[ht]
    \centering
    \begin{tabular}{|p{2cm}|p{2cm}|p{2cm}|}
        \hline
        \textbf{Total Elements} & \textbf{Proposed SHELBRS Scheme} & \textbf{Lyu et al. \cite{lyu2019privacy}}\\
        \hline
        10 & 0.024 & 2.79\\
        \hline
        20 & 0.083 & 5.48\\
        \hline
        40 & 0.302 & 11.13\\
        \hline
        80 & 1.149 & 22.32\\
        \hline
        100 & 1.933 & 28.03\\
        \hline
        1000 & 198.088 & 269.47\\ 
        \hline
    \end{tabular}
    \caption{Comparison of total execution time taken by proposed SHELBRS scheme and Lyu et al. \cite{lyu2019privacy}}
    \label{ComparingComputationCost}
\end{table}

\begin{figure}[ht]
\centering
\includegraphics[scale=0.45]{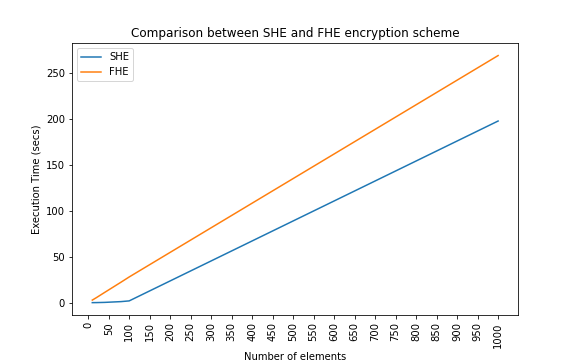}
\caption{Comparison of total execution time taken by proposed SHELBRS scheme and Lyu et al. \cite{lyu2019privacy}}
\label{fig:graph}
\end{figure}

\subsection{Security Analysis}
Our protocol aims to provide data confidentiality. It does not leak any meaningful information throughout the protocol. The security of the scheme is based on the following assumptions:
\begin{itemize}
    \item The ElGamal and Paillier schemes are secure.
    \item At least one of the servers is honest i.e. if one of the servers is malicious, the other server remains honest. 
    \item None of the servers collude.
\end{itemize}

Let us assume an Adversary $A$ plays the role of either a malicious server $Y$ or a malicious server $X$. Initially, $A$ is given the public keys and private key shares to perform paillier encryption $E^{+}$, elGamal encryption $E^{*}$, AddToMul and MulToAdd algorithms. $A$ is also given access to choose any arbitrary plaintext and can perform encryption to get the corresponding ciphertext. A user sends $E^{*}(m_{0})$ to the server $Y$ where $m_{0}$ is an integer except a value $0$. Now, A's goal is to find a challenge $m'_{0}$ such that $m'_{0} = m_{0}$. If A's goal is achieved, the security is broken. To prove $m_{0}'s$ security, given below are some lemmas.

\textbf{Lemma1}: If Server $Y$ is a malicious server and $A$ chooses to attack AddToMul algorithm, it has access to  $E^{*}(m_{0})$, $E^{+}(m_{0})$ and $E^{+}(E^{*}(m_{0}))$. Paillier encrypted $E^{+}(m_{0})$ and $E^{+}(E^{*}(m_{0}))$ terms are semantically secure and no decryption key $sk^{+}$ is associated with any server, so, no information regarding $m_{0}$ is leaked through these terms. Server $Y$ has key share $k_{1}^{*} = x_{1}$ and $E^{*}(m_{0})$ = $(m_{0}h^{r}, g^{r})$ where $r$ is randomly chosen from the group $Z_{N}^{*}$. The security of $E^{*}(m_{0})$ is based on SHE security proof\cite{lim2014faster} is perfectly secure. So, $A$ cannot learn anything about $m_{0}$ in the protocol. \\   

\textbf{Lemma2}: If Server $Y$ is a malicious server and $A$ chooses to attack MulToAdd algorithm, it has access to $k_{1}^{*} = x_{1}$. $E^{*}(m_{0})$, $E^{+}(E^{*}(m_{0}))$, $R' = g^{sx_{0}}$ and $c" = E^{+}(m_{0}h^{-s})$. $E^{*}(m_{0})$, $E^{+}(E^{*}(m_{0}))$ and $c"$ are secure according to lemma1. $A$ cannot learn about secret key share $x_{0}$ from $R'$ as $s$ is randomly chosen from the group $Z_{N}^{*}$. Therefore, the proposed scheme is secure against the malicious activity performed by server $Y$ itself.

Likewise, we can prove the data confidentiality using lemma1 and lemma2 when Server $X$ acts as an adversary $A$.

Now, we will handle the case when the client performs elGamal encryption on a message $m_{0}$ where $m_{0} = 0$. The ElGamal Encryption of $m_{0}$ plaintext results into one of the ciphertexts as "zero". This is not secure as it leaks information regarding plaintext data. To handle such problems, we can represent "zero" in the form
\begin{equation}
    \footnotesize
    \textbf{MulToAdd}(E^{+}(E^{*}(n_{1}))) * \textbf{MulToAdd}(E^{+}(E^{*}(n_{1})))^{-1} = E^{+}(0)
\end{equation}

\section{Conclusions and Future Work}
\label{sec:conclusions}
A lightweight privacy-preserving recommendation protocol for LBS was proposed in this paper. It incorporated Hilbert curve, collaborative filtering recommender based on co-occurrence matrix and SHE to recommend the services. Based on the simulation and experiments, we found that the computation cost for $5000$ POIs is $4355.921$ seconds. Compared with the state-of-the-art protocol, the proposed protocol takes less computation time and reduces complexity, providing at par security.  
As the future direction of the work, we would like to extend our protocol for the larger geographical area as we focused herein only on item-based filtering on a single geographical area.

\bibliography{references.bib}

\begin{thebibliography}{10}

\bibitem{Strava}
Mark~Ingle Rhys~Fenwick, Mike~Hittle and Oliver White.
\newblock Fitness app strava lights up staff at military bases.
  https://www.bbc.com/news/technology-42853072.
\newblock {\em BBC Journal Archive}, 2018.

\bibitem{sagan2012space}
Hans Sagan.
\newblock {\em Space-filling curves}.
\newblock Springer Science \& Business Media, 2012.

\bibitem{sarwar2001item}
Badrul Sarwar, George Karypis, Joseph Konstan, and John Riedl.
\newblock Item-based collaborative filtering recommendation algorithms.
\newblock In {\em Proceedings of the 10th international conference on World
  Wide Web}, pages 285--295, 2001.

\bibitem{moon2001analysis}
Bongki Moon, Hosagrahar~V Jagadish, Christos Faloutsos, and Joel~H. Saltz.
\newblock Analysis of the clustering properties of the hilbert space-filling
  curve.
\newblock {\em IEEE Transactions on knowledge and data engineering},
  13(1):124--141, 2001.

\bibitem{brakerski2014leveled}
Zvika Brakerski, Craig Gentry, and Vinod Vaikuntanathan.
\newblock (leveled) fully homomorphic encryption without bootstrapping.
\newblock {\em ACM Transactions on Computation Theory (TOCT)}, 6(3):1--36,
  2014.

\bibitem{lyu2019privacy}
Qiuyi Lyu, Yu~Ishimaki, and Hayato Yamana.
\newblock Privacy-preserving recommendation for location-based services.
\newblock In {\em 2019 IEEE 4th International Conference on Big Data Analytics
  (ICBDA)}, pages 98--105. IEEE, 2019.

\bibitem{gentry2009fully}
Craig Gentry.
\newblock Fully homomorphic encryption using ideal lattices.
\newblock In {\em Proceedings of the forty-first annual ACM symposium on Theory
  of computing}, pages 169--178, 2009.

\bibitem{gruteser2003anonymous}
Marco Gruteser and Dirk Grunwald.
\newblock Anonymous usage of location-based services through spatial and
  temporal cloaking.
\newblock In {\em Proceedings of the 1st international conference on Mobile
  systems, applications and services}, pages 31--42, 2003.

\bibitem{melchor2008fast}
Carlos~Aguilar Melchor and Philippe Gaborit.
\newblock A fast private information retrieval protocol.
\newblock In {\em 2008 IEEE International Symposium on Information Theory},
  pages 1848--1852. IEEE, 2008.

\bibitem{lien2013novel}
I-Ting Lien, Yu-Hsun Lin, Jyh-Ren Shieh, and Ja-Ling Wu.
\newblock A novel privacy preserving location-based service protocol with
  secret circular shift for k-nn search.
\newblock {\em IEEE Transactions on Information Forensics and Security},
  8(6):863--873, 2013.

\bibitem{utsunomiya2015lpcqp}
Yasuhito Utsunomiya, Kentaroh Toyoda, and Iwao Sasase.
\newblock Lpcqp: Lightweight private circular query protocol for
  privacy-preserving k-nn search.
\newblock In {\em 2015 12th Annual IEEE Consumer Communications and Networking
  Conference (CCNC)}, pages 59--64. IEEE, 2015.

\bibitem{pingley2012context}
Aniket Pingley, Wei Yu, Nan Zhang, Xinwen Fu, and Wei Zhao.
\newblock A context-aware scheme for privacy-preserving location-based
  services.
\newblock {\em Computer Networks}, 56(11):2551--2568, 2012.

\bibitem{sun2019towards}
Gang Sun, Liangjun Song, Dan Liao, Hongfang Yu, and Victor Chang.
\newblock Towards privacy preservation for “check-in” services in
  location-based social networks.
\newblock {\em Information Sciences}, 481:616--634, 2019.

\bibitem{lu2012recommender}
Linyuan L{\"u}, Mat{\'u}{\v{s}} Medo, Chi~Ho Yeung, Yi-Cheng Zhang, Zi-Ke
  Zhang, and Tao Zhou.
\newblock Recommender systems.
\newblock {\em Physics reports}, 519(1):1--49, 2012.

\bibitem{badsha2016practical}
Shahriar Badsha, Xun Yi, and Ibrahim Khalil.
\newblock A practical privacy-preserving recommender system.
\newblock {\em Data Science and Engineering}, 1(3):161--177, 2016.

\bibitem{zhang2018fm}
Xiaojian Zhang, Limin Yu, Minjuan Wang, and Wanlin Gao.
\newblock Fm-based: Algorithm research on rural tourism recommendation
  combining seasonal and distribution features.
\newblock {\em Pattern Recognition Letters}, 2018.

\bibitem{horowitz2018eventaware}
Daniel Horowitz, David Contreras, and Maria Salam{\'o}.
\newblock Eventaware: A mobile recommender system for events.
\newblock {\em Pattern Recognition Letters}, 105:121--134, 2018.

\bibitem{qi2019time}
Lianyong Qi, Ruili Wang, Chunhua Hu, Shancang Li, Qiang He, and Xiaolong Xu.
\newblock Time-aware distributed service recommendation with
  privacy-preservation.
\newblock {\em Information Sciences}, 480:354--364, 2019.

\bibitem{papakyriakopoulos2020political}
Orestis Papakyriakopoulos, Juan Carlos~Medina Serrano, and Simon Hegelich.
\newblock Political communication on social media: A tale of hyperactive users
  and bias in recommender systems.
\newblock {\em Online Social Networks and Media}, 15:100058, 2020.

\bibitem{lim2014faster}
Hoon~Wei Lim, Shruti Tople, Prateek Saxena, and Ee-Chien Chang.
\newblock Faster secure arithmetic computation using switchable homomorphic
  encryption.
\newblock {\em IACR Cryptol. ePrint Arch.}, 2014:539, 2014.

\end{thebibliography}

\end{document}